\documentclass{aastex}

\newcommand{\taub}{{\mbox{\boldmath $\tau$}}}
\newcommand{\hb}{{\bf h}}
\newcommand{\kb}{{\bf k}}
\newcommand{\rb}{{\bf r}}
\newcommand{\tb}{{\bf t}}
\newcommand{\uub}{{\bf u}}
\newcommand{\xb}{{\bf x}}

\shorttitle{Non-Gaussian Random Fields}
\shortauthors{Vio, Wamsteker}

\begin{document}

\title{Numerical Simulation of
Non-Gaussian Random Fields \\ with Prescribed Correlation Structure}
\author{Roberto Vio\altaffilmark{1}}
\affil{Chip Computers Consulting s.r.l., Viale Don L.~Sturzo 82, \\
S.Liberale di Marcon, 30020 Venice, Italy}
\email{robertovio@tin.it}
\and
\author{Paola Andreani\altaffilmark{2}}
\affil{Osservatorio Astronomico di Padova, vicolo dell'Osservatorio 5,
35122 Padua, Italy}
\email{andreani@mpe.mpg.de}
\and
\author{Willem Wamsteker}
\affil{ESA IUE Observatory, Apartado 50727, 28080 Madrid, Spain}
\email{ww@vilspa.esa.es}
\altaffiltext{1}{ESA IUE Observatory, Apartado 50727, 28080 Madrid,
Spain}
\altaffiltext{2}{Max-Planck Institut f\"ur Extraterrestrische Physik,
Postfach 1312
85741 Garching, Germany}

\begin{abstract}

In this paper we will consider the problem of the numerical simulation
of non-Gaussian, scalar random fields with a prescribed correlation
structure provided either by a theoretical model or computed on a set
of observational data.
Although, the numerical generation of a generic,
non-Gaussian random field is a trivial operation, the task becomes
tough when constraining the field with a prefixed correlation structure.
At this regards, three numerical methods, useful for astronomical
applications, are presented. The limits and capabilities of each method
are discussed and the pseudo-codes describing the numerical
implementation
are provided for two of them.

\end{abstract}


\section{INTRODUCTION}

Computer-aided modeling is becoming an essential tool in designing
new experiments and in testing theoretical models against the
observational data. For example, because of the cost of any space-based
telescope, nowadays it is not even conceivable to plan a mission without
first simulating the performances either of the instruments and/or of
the observing mode. It is obvious to stress that the reliability of such
simulations depends critically on the possibility to reproduce realistic
physical scenarios.

A wide assumption, which is often made because of its simplicity, is
that the processes underlying a given physical phenomenon
obey Gaussian, and therefore linear, statistics. However, although as
practical as this assumption could be, it is not applicable for most of
the physical systems which, on the contrary, are expected to be
characterized by nonlinear behaviours.
Some examples:
\begin{itemize}
\item[-] the high spatial resolution observations of sky images are
revealing a lot of details of the sky emission which are not of easy
interpretation (e.g. \citet{her98}).
For istance, in the far infrared spectral domain, the studies of source
properties imply the disentangling of the source emission
and position from those of a much stronger background
whose spatial structure is highly non-Gaussian.
This task becomes crucial for most space-borne surveys of the
extragalactic sky and of star-forming regions in the Galaxy
for which it is of interest to simulate
the far-IR and sub-mm emission of the
Galaxy and the source confusion in the beam. Their nature is
intrinsically non-Gaussian and to match their observed
properties an appropriate method must be used.
\item[-]
The interpretation of the new flow of data on the Cosmic Microwave
Background
(CMB) spatial distribution, from present and future experiments
(BOOMERANG, MAXIMA, MAP and PLANCK), is involving a large theoretical
effort (see e.g. \citet{ver00,mat00,con01} and references therein).
Studies of the spatial structure of the CMB provide fundamental clue on
the
physical processes generating the primordial density fluctuations
which are thought to be at the origin of the present-day structures.
There are deep theoretical motivations, both in the framework of
inflationary
models and in cosmological defects scenarios, to consider
the initial density perturbations obeying non-Gaussian statistics.
In any case, the subsequent growth of these density fluctuations,
triggered by
the gravitational potential, makes the late time evolution nonlinear.
The testing of these predictions requires algorithms able
to discerne the true statistical nature of the observed fields.
Some of their properties can be analytically recovered
but the bulk of
non-Gaussian random fields characteristics can be inferred
only through simulations (see e.g., \citet{mos91}).
\end{itemize}

The aim of this paper is to provide a general and mathematical approach
to the problem of generating non-Gaussian random fields
when a correlation
structure is given either by a theoretical model or by the
statistical analysis of experimental data.
Some numerical procedures to perform simulations of such fields are also
presented.
The arguments are outlined on a quite general basis in view of
applications
in a wide astrophysical context.
The reason to fix the correlation structure
is that this is the simplest way to obtain non-trivial (i.e. non-pure
noise) fields (see below).

The problem can be formalized as follows. A real, random field $R(\tb)$
can be defined as a collection of random variables $\{ {\bf r} \}$ at
points with coordinates $\{ \tb \}= \{(t_1,t_2,\ldots,t_n) \}$
\footnote{From now on, in order to distinguish them from scalar
quantities, we will denote vector quantities in
boldface.} in a {\it n}-th dimensional ``parameter space''. In other
words, for each ``position'' $\hat \tb$, $R(\hat \tb)= {\bf r}$, where
${\bf r}$ is a vector characterized by a multidimensional distribution
function $F_R({\bf r})$ and a multidimensional probability density
function $f_R({\bf r})$. According to the particular problem at hand,
$\{ \tb \}$ may correspond to a set of spatial/angular coordinates
(spatial random fields), to time (time processes), to a mix of these two
(spatio-temporal random fields) or even to more general situations.

In many practical situations they are of interest the so called {\it
scalar} random fields, where ${\bf r} \equiv r$. This means that for a
specific $\hat \tb$, $R(\hat \tb)$ is characterized by a scalar
random variable $r$ with one-dimensional distribution function (DF) $F_R(r)$ 
and one-dimensional
probability density function (PDF) $f_R(r)$. In this paper we will
consider only this kind of random fields. The more general case of
vector-valued random fields represents a more complex problem and
will not be addressed in this work (for more details about this topic
see \citet{pop98}).

The main problem in the numerical simulation of a generic $R(\tb)$ is
that, in general, given two arbitrary ``positions'', say $\tb_1$ and
$\tb_2$, $R(\tb_1)$ and $R(\tb_2)$ are not independent one of the
other. As well known from the standard theory of random processes (e.g.
\citet{gri95}), the practical applications to generate random fields
requires to put some constraints. The most common choice is that,
specified
the distribution function $F_R(r)$, $R(\tb)$ be completely characterized
by the covariance function, $\xi_R(\tb_1,\tb_2)$, with
\begin{equation} \label{eq:covariance}
\xi_R(\tb_1,\tb_2)= {\rm E} \left[ R(\tb_1) ~R(\tb_2) \right],
\end{equation}
where ${\rm E}[.]$ stands for expected value. The reason is that
$\xi_R(\tb_1,\tb_2)$ represents the simplest form of mutual relationship
between the elements of $R(\tb)$.

In astronomical applications, very often it is possible to adopt some
simplifying conditions.
In particular, it is possible to assume that $R(\tb)$ is isotropic. This
means that the covariance function depends on the length of the
vector $\tb_1-\tb_2$ but not on its direction: $\xi_R(\tb_1,\tb_2)=
\xi_R(\Vert \tb_1-\tb_2 \Vert)$ \footnote{We remind that for a column
vector ${\bf a}=(a_1,a_2,\ldots, a_N)^T$, $\Vert {\bf a} \Vert= [
{\bf a}^T {\bf a}]^{1/2}= (\sum_{i=1}^N a_i^2)^{1/2}$ provides its
length (norm). Here ${\bf a}^T$ means the transpose of vector ${\bf
a}$.}. In other words, $R(\tb)$ is characterized by a spherical
symmetry. This property is very useful since it allows to characterize
$R(\tb)$ through the correlation function
\begin{equation} \label{eq:rho}
\rho_R(\tau)= {\rm E} \left[ \frac{(R(\tb) - \mu_R) ~(R(\tb + \taub)
- \mu_R)}{\sigma_R^2} \right],
\end{equation}
where $\tau= \Vert \taub \Vert$, and $\mu_R$ and $\sigma^2_R$ are,
respectively, the mean and the variance corresponding to the
distribution $F_R(r)$.

Although the isotropic case is of large interest in astronomical
applications, here we prefer to adopt a more general formalism,
suited for all
homogeneous fields. Indeed, in
this case, the covariance function depends on $\tb_1-\tb_2$. According
to this definition, $\rho(\tau)$, in equation (\ref{eq:rho}) has to be
replaced by
$\rho(\taub)$.

\section{PRELIMINARY NOTES} \label{sec:notes}

Most of the techniques for simulating a non-Gaussian, scalar, random
field $R(\tb)$, with a prescribed correlation function, $\rho_R(\taub)$,
and a prescribed one-dimensional marginal $F_R(r)$, are more or less
explicitly based on the two following steps:
\begin{itemize}
\item[-] generation of a zero-mean, unit-variance, scalar, Gaussian
random field $X(\tb)$ with a prefixed correlation structure
$\rho_X(\taub)$;
\item[-] mapping (transformation) $X(\tb) \rightarrow R(\tb)$ according
to
\begin{equation} \label{eq:map}
R(\tb)= g [ X(\tb) ],
\end{equation}
where $g[.]$ represents an appropriate function. This operation is named
{\it memoryless transformation} since the value of $R(\tb)$ at an
arbitrary $\hat \tb$ depends only on the value of $X(\hat \tb)$.
\end{itemize}
The rationale behind such a procedure is that the direct generation of a
generic $R(\tb)$, with a specific $\rho_r(\taub)$, is a very difficult
operation. The Gaussian case represents a useful exception. Hence, it
results much easier to obtain $R(\tb)$ by transforming a precomputed
$X(\tb)$. However, after the mapping (\ref{eq:map}), in general
$\rho_X(\taub)$ does not coincide with $\rho_R(\taub)$. Therefore, it is
necessary to transform an $X(\tb)$ characterized by an appropriate
$\rho_X(\taub)$ whose functional form depends on $\rho_R(\taub)$.

It is well known from elementary statistics that for a mapping
$r=g(x)$, with $x$ the standard one-dimensional Gaussian variable, the
PDF of the random variable $r$ can be obtained from that of the random
variable $x$ via a change of variable technique. In the general case
that the tranformation $g(.)$ is not one-to-one, if the equation
\begin{equation} \label{eq:transform}
g(x)-r=0
\end{equation}
has a numerable set of $M$ real solutions $\{x_1(r), x_2(r), \ldots,
x_M(r)\}$, and if $g_j'=[d g(x) / d x]_{x=x_j}$, $j=1,2, \ldots,
M$ exist, then $f_R(r)$ is given by \citep{pap91}
\begin{equation} \label{eq:change1}
f_R(r)= \frac{1}{\sqrt{2 \pi}} \sum_{j=1}^{M} \frac{{\rm
e}^{-x_j^2(r)/2}}{\left| g_j' \right|}.
\end{equation}
In correspondence to the values $r^*$, where equation
(\ref{eq:transform}) does not have real solutions, it happens that
$f_R(r^*)=0$. Furthermore, the correlation function $\rho_R(\taub)$ is
given by \citep{gri95}
\begin{equation} \label{eq:change}
\rho_R(\taub)= \frac{1}{\sigma_R^2} \int_{-\infty}^{\infty}
\int_{-\infty}^{\infty} [g(x_1) - \mu_R] ~[g(x_2) - \mu_R] ~\phi(x_1,
x_2; \rho_X(\taub))~dx_1 ~dx_2,
\end{equation}
where, $x_1= x(\tb)$ and $x_2= x(\tb+\taub)$, and
\begin{equation}
\phi(x_1, x_2; \rho_X(\taub))= \frac{1}{2 \pi (1
-\rho_X^2(\taub))^{1/2}} ~\exp \left(
- \frac{x_1^2 + x_2^2 - 2 \rho_X(\taub) x_1 x_2}{2 (1 -
\rho_X^2(\taub))} \right).
\end{equation}

At first sight, from these equations it may seem that, given the
appropriate function $g(.)$ and the covariance function $\rho_X(\taub)$,
obtained via the inversion of equation (\ref{eq:change}), it is possible
to generate an $R(\tb)$ characterized by an arbitrary $\rho_R(\taub)$.
In reality, given a generic $g(.)$, there is no guarantee that equation
(\ref{eq:change}) can be inverted. Furthermore, it is possible to show
\citep{ogo96} that $\rho_R(\taub)$ can take values only in the interval
\begin{equation} \label{eq:interval}
\rho_R(\taub) \in [\rho^*,1],
\end{equation}
where
\begin{equation} \label{eq:limit}
\rho^*=\frac{1}{\sigma_R^2} \left( \int_0^1 F_R^{-1}(\alpha)
~F_R^{-1}(1-\alpha) ~d\alpha - \mu_R^2 \right),
\end{equation}
with
\begin{equation}
F_R^{-1}(\alpha)=\inf\{r:F_R(r) > \alpha \}
\end{equation}
providing the smallest value of the random variable $r$ satisfying the
condition that $F_R(r) > \alpha$. In particular, $\rho^*=-1$ only for
symmetric distributions. For example, $\rho^* \simeq -0.645$ in case of
the exponential PDF: $f_R(r)= \beta ~\exp(-\beta r), r \ge 0$. This
shows that, in general, for a fixed $g(.)$ it is not possible to obtain
a $\rho_R^*(\taub)$ presenting values external to the interval
(\ref{eq:interval}).

Another problem stems from the fact that the function $\rho_X(\taub)$,
necessary to obtain the target $\rho_R(\taub)$ via equation
(\ref{eq:change}), must be a non-negative definite function \footnote{It
should be remembered that only for a non-negative defined function
the corresponding Fourier transform has non-negative values.
Therefore, $\rho_X(\taub)$ must share this property since, according to
the Wiener-Khinchin theorem, the power-spectrum of a process is provided
by the Fourier transform of the corresponding correlation function.}
otherwise the generation of the aimed $R(\tb)$ is prevented since
$\rho_X(\taub)$ does not represent any correlation function.

In principle, equation (\ref{eq:change}) can be used to obtain
$\rho_X(\taub)$ in a closed form, but in reality such an approach can be
followed only in a limited number of cases. Indeed, very often the
transformation (\ref{eq:change}) has a very complex form and can be
inverted only via a numerical approach.

\section{ANALYTICAL METHOD}

\begin{figure}[h]         
\plotone{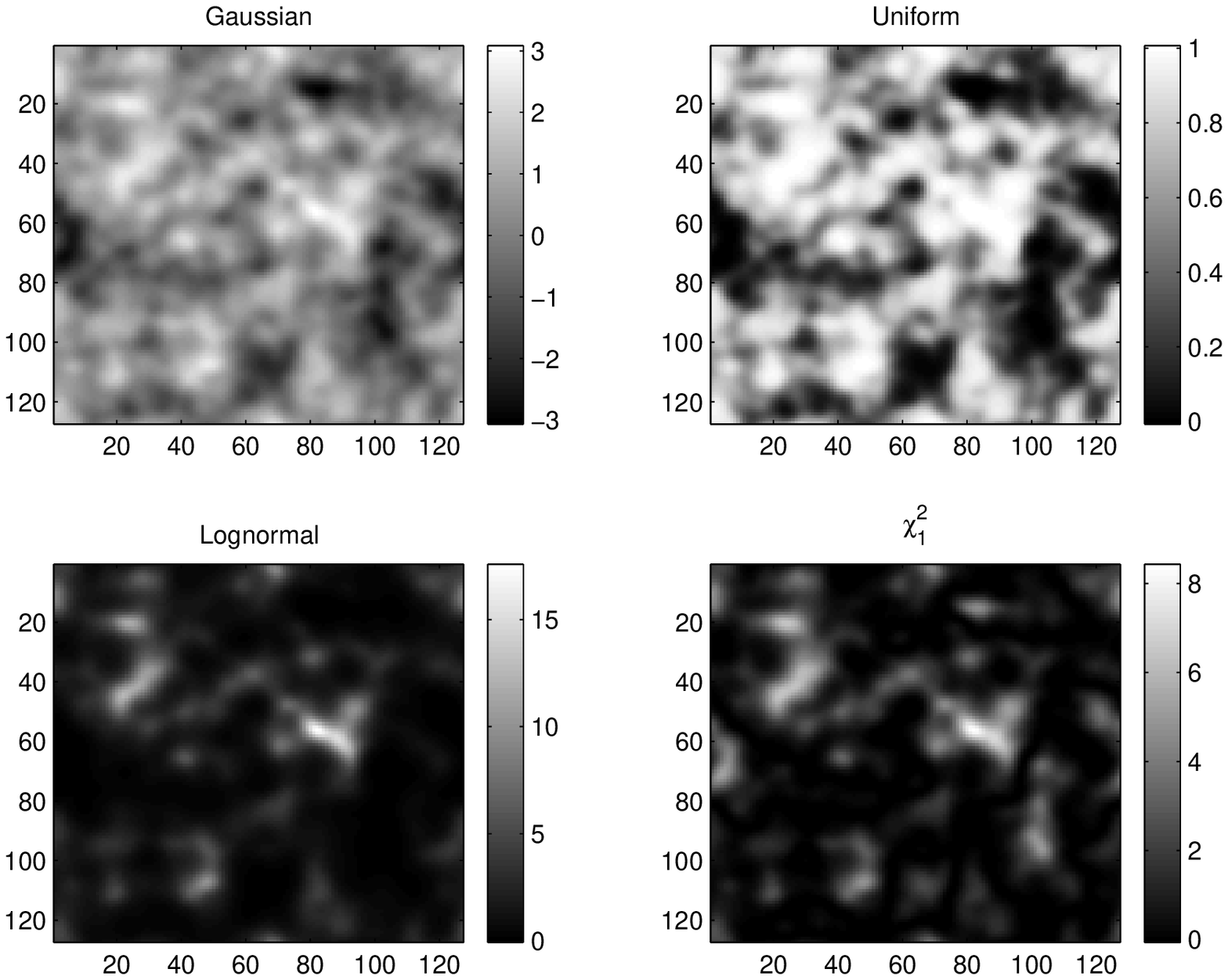}
\caption{\label{fig:fields} Examples of non-Gaussian random fields characterized
by the same correlation function.}
\end{figure}

Certainly one of the most effective method for generating $R(\tb)$ is
represented the analytical handling of equations (\ref{eq:change1}) and
(\ref{eq:change}). Unfortunately, this is also the most difficult
approach to pursue; only in a limited number of cases it has been
possible to find out the analytical relationship between $\rho_X(\taub)$
and $\rho_R(\taub)$. Some useful examples are presented below (see also
figure \ref{fig:fields}):

\subsection{Lognormal Fields}

If $X(\tb)$ is a homogeneous, zero-mean, unit-variance Gaussian fields
with a correlation function $\rho_X(\taub)$, the random fields obtained
via
\begin{equation}
L(\tb)=e^{\mu+\sigma X(\tb)}
\end{equation}
are called Lognormal fields since they are characterized by the
one-dimensional marginal Lognormal PDF
\begin{equation}
f_L(l)=\frac{1}{l ~\sigma \sqrt{2 \pi}}~ e^{- \left( \ln l - \mu
\right)^2/(2 \sigma^2)}, \qquad l > 0.
\end{equation}
It is possible to show \citep{van84} that the moments of order $k$ of
$L(\tb)$ are given by
\begin{equation}
E[L^k]= e^{k \mu + k^2 \sigma^2 /2}.
\end{equation}
In particular, the mean and the variance are
\begin{equation} \label{eq:momL}
\mu_L= e^{\mu+\sigma^2/2}, \qquad \sigma^2_L= e^{2 \mu
+\sigma^2}(e^{\sigma^2}-1),
\end{equation}
respectively. Furthermore, it is possible to show that the relationship
between $\rho_L(\taub)$ and $\rho_X(\taub)$ can be expressed in the form

\begin{equation} \label{eq:corrL}
\rho_L(\taub)= \frac{e^{\sigma^2 \rho_X(\taub)} - 1}{e^{\sigma^2}-1}.
\end{equation}
From this equation it is trivial to see that, when $\sigma= 1$,
the lower bound of $\rho_L(\taub)$ is $\simeq -0.368$.

\subsection{Gamma Fields}

If $X_s(\tb)$, $s=1,2,\ldots, 2 m$, is a collection of independent
zero-mean, unit-variance Gaussian fields with the same correlation
function $\rho_X(\taub)$, the random fields obtained via
\begin{equation} \label{eq:sumG}
G_{m}(\tb)=\frac{1}{2} \sum_{s=1}^{2 m} X_s^2(\tb)
\end{equation}
are called Gamma fields. That is because the corresponding
one-dimensional marginal PDF is a Gamma distribution with $m$ degrees of
freedom
\begin{equation}
f_{G_{m}}(g)= \frac{1}{\Gamma(m)} ~ g^{m-1} e^{-g}, \qquad g \geq 0,
\end{equation}
where $\Gamma(.)$ is the Gamma function.

\noindent
It can be shown that the moments of order $k$ of $G_m$ are given by
\begin{equation}
E[G_m^k]= \frac{\Gamma(m+k)}{\Gamma(m)}= \prod_{i=0}^{k-1} (m +
i), \qquad k > -m.
\end{equation}
In particular, the mean and the variance are
\begin{equation} \label{eq:momG}
\mu_{G_m}= \sigma^2_{G_m}= m.
\end{equation}
It can also be shown \citep{has98} that, idependently from the value of
$m$, the relationship between $\rho_{G_m}(\taub)$ and $\rho_X(\taub)$
can be expressed in the form
\begin{equation} \label{eq:corrG}
\rho_{G_m}(\taub)= \rho_X^2(\taub).
\end{equation}
The class of the Gamma fields is interesting since it contains, as
particular cases, both the Chi-Square and the Exponential fields.

From equation (\ref{eq:corrG}) it appears that the lower bound of 
$\rho_{G_m}(\taub)$ is zero. In
other words, through the mapping (\ref{eq:sumG}) it is not possible to
obtain $G_m(\tb)$ characterized by correlation functions with negative
values. Here, however, it is necessary to stress that such a limit is
not intrinsic to the Gamma fields, but only to the transformation
(\ref{eq:sumG}). In fact, through different mappings it is possible to
generate $G_m(\tb)$ with correlation functions having negative values
(see below).

\subsection{Beta Fields}

Given two independent Gamma fields, say $G_m(\tb)$ and $G_n(\tb)$,
characterized by the same correlation function $\rho_G(\taub)$, the
random fields obtained via
\begin{equation} \label{eq:beta}
B_{mn}(\tb)= \frac{G_m(\tb)}{G_m(\tb) + G_n(\tb)}
\end{equation}
are called Beta fields because their one-dimensional marginal PDF is a
Beta$(m,n)$ distribution
\begin{equation}
f_{B_{mn}}(b)= \frac{1}{B(m,n)}~ x^{m-1} (1-x)^{n-1}, \qquad 0 \leq b
\leq 1.
\end{equation}
It can be shown that the moments of order $k$ of $B_{mn}$ are given by
\begin{equation}
E[B^k]= \frac{\Gamma(m+k) \Gamma(m+n)}{\Gamma(m) \Gamma(m+n+k)}.
\end{equation}
In particular, the mean and the variance are
\begin{equation} \label{eq:momB}
\mu_{B_{mn}}= \frac{m}{m+n}, \qquad \sigma^2_{B_{mn}}=
\frac{mn}{(m+n)^2(m+n+1)},
\end{equation}
respectively.

It can be also shown \citep{has98} that the relationship between
$\rho_{B_{mn}}(\taub)$ and $\rho_X(\taub)$ can be expressed in the form
\begin{equation} \label{eq:corrB}
\rho_{B_{mn}}(\taub)= 1 - S_{m+n}[\rho_X(\taub)], \qquad n+m > 1,
\end{equation}
where
\begin{equation}
S_q(\rho)= q \left( \frac{1-\rho}{-\rho} \right)^q \left[\log(1-\rho)
- \sum_{i=1}^{q-1} \frac{1}{i}
\left( \frac{-\rho}{1-\rho} \right)^i \right], \qquad 0 \leq \rho \leq
1, ~q\in\{1,2,\ldots\},
\end{equation}
with the end values $S_q(0)= 1$ and $S_q(1)=0$.
The class of the Beta fields is basic for describing variables bounded
at both sides. For example, $B_{11}(\tb)$ corresponds to the Uniform
field.

As well as for the Gamma fields, also for the Beta fields obtained
through mapping (\ref{eq:beta}) it happens that the lower bound of 
$\rho_{B_{mn}}(\taub)$ is zero. Again, this limit is not intrinsic to 
the Beta fields but only to the particular mapping used.

\section{NUMERICAL METHOD}

In case one is interested in a $R(\tb)$ characterized by an $F_R(r)$ not
reproducible through a function $g(.)$ available in analytical form
and/or that does not permit an easy calculation of $\rho_X(\taub)$, it
is necessary to resort to numerical methods. Regarding this, the
following presents three methods that can be useful in astronomical
applications.

\subsection{Change of Variable Method} \label{sec:inversion}

The most obvious method is based on the numerical inversion of equation
(\ref{eq:change}). Whenever possible, this is the 'method' to use since,
contrary to the procedures presented below, it is able to
provide exact results (within the limits of the numerical computation).
In particular, this inversion  operation is feasible when $g(.)$ is a
monotonic increasing, real function. Indeed, in this case the
relationship between $\rho_R(\taub)$ and $\rho_X(\taub)$ can always be
inverted. 

In the case of distributions $F_r$ with no atoms (a concentration of a
finite probability mass at a point), a very useful kind of monotonic
increasing functions $g(.)$ is represented by the mapping
\begin{equation} \label{eq:map1}
R(\tb)= F^{-1}_R \{ F_X [ X(\tb) ] \},
\end{equation}
where $F_X$ denotes the Gaussian distribution function and $F^{-1}_R$
the inverse distribution function of $R(\tb)$. Indeed, through
$g(.)=F^{-1}_R \{ F_X(.) \}$ the generation of fields $R(\tb)$, with
arbitrary one-dimensional marginal distribution functions, is possible.
Furthermore, it can be shown \citep{ogo96, gri95} that via the mapping (\ref{eq:map1})
it is possible to obtain $R(\tb)$ characterized by $\rho_R(\taub)$
fully exploiting the interval (\ref{eq:interval}) with $\rho^*$ that can be
simplified to the form
\begin{equation}
\rho^*= \frac{{\rm E}[g(x)~g(-x)] - \mu_R^2}{\sigma_R^2}.
\end{equation}

\begin{figure}[h]          
\plotone{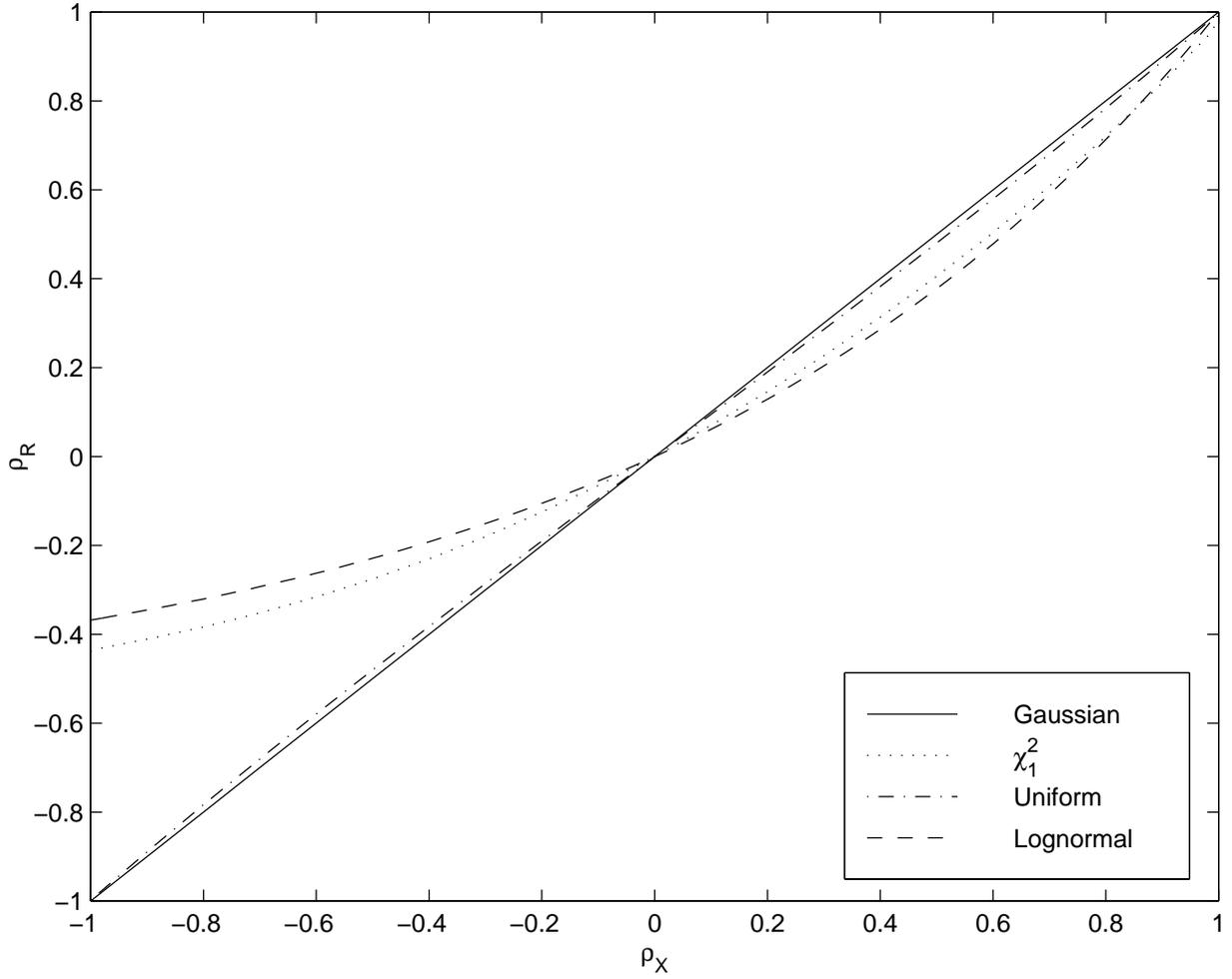}
\caption{\label{fig:rho} Relationship between the correlation
function $\rho_R(\tau)$ of some non-Gaussian random fields obtained via
transformation (\ref{eq:map1}) and the correlation function
$\rho_X(\tau)$ of the Gaussian fields used in such transformation.}
\end{figure}

Figure \ref{fig:rho} shows the relationship between $\rho_X$ and
the $\rho_R$, concerning some well known $F_R$, obtained via
equations (\ref{eq:map1}) and (\ref{eq:change}). From this figure it is
possible to realize some interesting points that can also be proved via
theoretical arguments \citep{gri95, gri98}:
\begin{itemize}
\item[-] $\rho_R(\taub)$ is an increasing function of $\rho_X(\taub)$;
\item[-] $| \rho_R(\taub) | \leq | \rho_X(\taub) |$;
\item[-] the difference between $\rho_R(\taub)$ and $\rho_X(\taub)$ are
not significant for a broad range of values of these functions.
\end{itemize}
As explained in Section \ref{sec:notes}, once $\rho_X(\taub)$ has been 
calculated, it is necessary to check that this function is 
non-negative definite. A possibility consists in evaluating the
Fourier transform of $\rho_X(\taub)$ and verifying that it
presents no negative values.

The only concern regarding the numerical inversion of equation
(\ref{eq:change}) is that, in general, this operation requires 
the calculation  of a large number of double integrals. 
In certain situations, that could
represent a computationally too expensive problem, and therefore it is
necessary to resort to other numerical techniques.

\subsection{Hermite Expansion Method}

An alternative approach for the simulation of a non-Gaussian $R(\tb)$ is
based on the expansion of the field in Hermite polynomials\footnote{
This expansion, known also as Edgeworth expansion,
was already applied in a Cosmological context
(see \citet{col94})}. These
polynomials can be defined through the Rodriguez's formula
\begin{equation}
{\rm H}_n(x)= (-1)^n e^{x^2/2} \frac{d^n}{d x^n} e^{-x^2/2}, \qquad
n=0,1,\ldots,
\end{equation}
and have the important property of being orthogonal relative to the
standard Gaussian distribution, so that
\begin{equation} \label{eq:hnormal}
\int_{-\infty}^{+\infty} {\rm H}_m(x) {\rm H}_n(x)~ \frac{1}{\sqrt{2
\pi}} ~e^{-x^2/2} dx= n! ~\delta_{mn}
\end{equation}
where $\delta_{mn}$ is the Kronecker function.

\noindent
An explicit expression for ${\rm H}_n(x)$ is given by \citet{bli98}
\begin{equation} \label{eq:expherm}
{\rm H}_n(x)= n! \sum_{k=0}^{\lfloor n/2 \rfloor} \frac{(-1)^k
x^{n-2k}}{k! (n-2k)! ~2^k},
\end{equation}
where $\lfloor z \rfloor$ means the largest integer $ k \leq z$.

\noindent
A field $R(\tb)$ can be expanded according to
\begin{equation} \label{eq:hexpansion}
R^*(\tb)= \sum_{k=0}^{N_H} a_k {\rm H}_k (X(\tb)),
\end{equation}
where the coefficients $\{ a_k \}$ are unknown and must be determined.
In the practical applications, also the ``optimal'' value $N_H$ has to
be determined.

\begin{figure}[h]          
\plotone{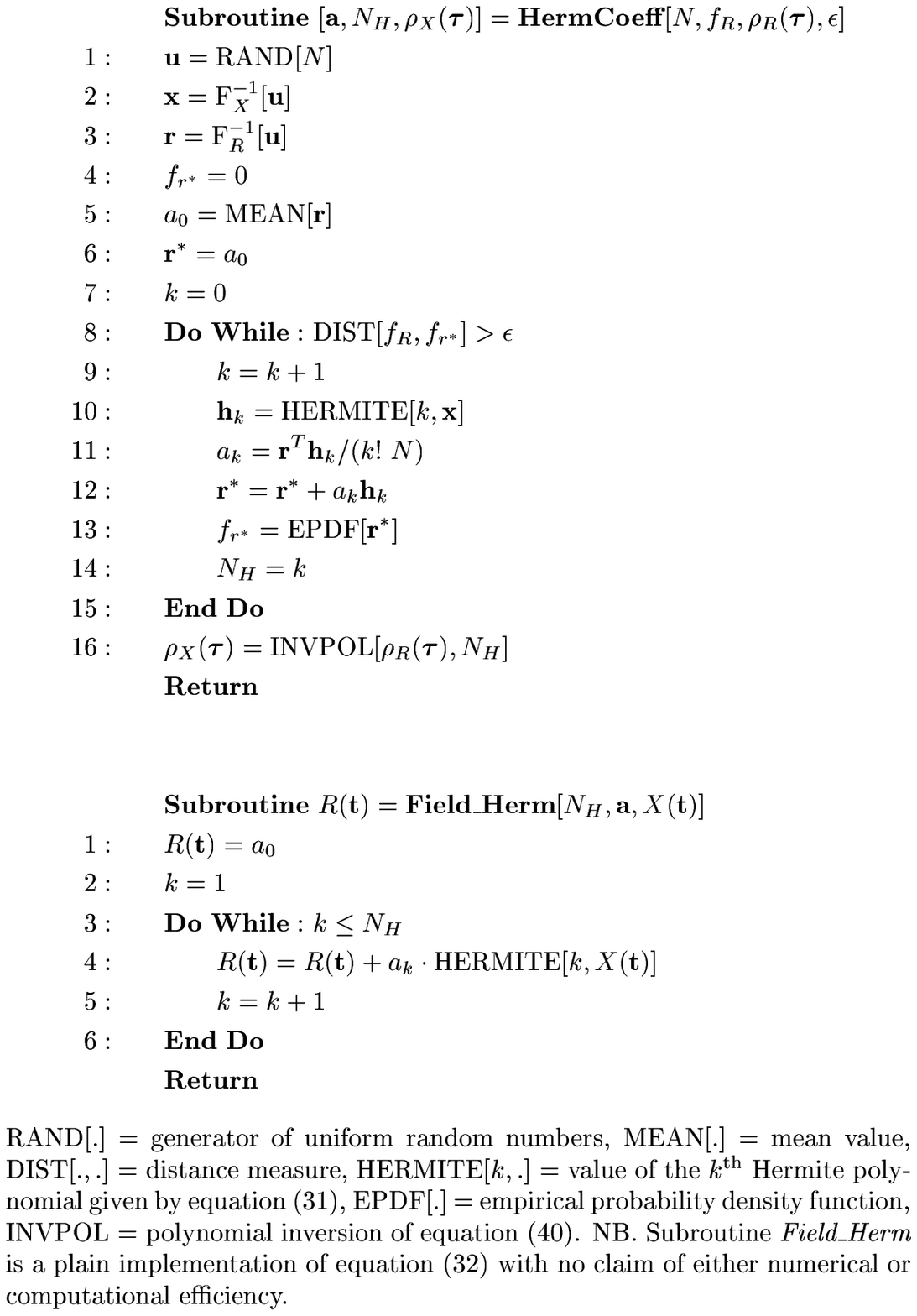}
\caption{\label{fig:code2} Subroutines {\it HermCoeff} and {\it
Field\_Herm}.}
\end{figure}

One possibility for obtaining the coefficients $\{ a_k \}$, for a fixed
$N_H$, is to minimize the objective function \citep{gri95}
\begin{equation} \label{eq:criterion}
\delta^2={\rm E} [|R(\tb) - R^*(\tb)|^2 ]
\end{equation}
that yields the conditions
\begin{equation}
{\rm E} \left[ \left( R(\tb) - \sum_{l=0}^{N_H} a_l ~{\rm H}_l(X(\tb))
\right) {\rm H}_k (X(\tb)) \right]= 0, \qquad k=0,1, \ldots N_H,
\end{equation}
so that
\begin{equation}
a_k= \frac{1}{k!} ~{\rm E}[R(\tb) ~{\rm H}_k(X(\tb))], \qquad
k=0,1,\ldots, N_H,
\end{equation}
because of equation (\ref{eq:hnormal}). Here, the important point is
that the coefficients $\{ a_k \}$ are independent from the structure of
$X(\tb)$ since, for a specific position $\hat \tb$, the value of $R(\hat
\tb)$ depends only on $X(\hat \tb)$. That allows us to estimate such
coefficients by means of
\begin{equation}
a_k=\frac{1}{k!} ~{\rm E}[r ~{\rm H}_k(x)],
\end{equation}
where $x$ is the standard Gaussian random variable, and $r$ is the
random variable distribuited according to the marginal distribution
required for $R(\tb)$.  Following this approach, the procedure
implemented in the subroutine {\it HermCoeff} in figure \ref{fig:code2}
is:
\begin{enumerate}
\item generation of a large (column) array of independent and uniform
random deviates $\uub= [ u_1, u_2, \ldots, u_N ]^T$;
\item mapping of $\uub$ in two arrays $\xb= [x_1,x_2,\ldots,
x_N]^T=F_X^{-1}(\uub)$ and $\rb= [r_1,r_2,\ldots, r_N]^T=
F_R^{-1}(\uub)$. Here, $\xb$ is an array of independent standard
Gaussian random deviates, whereas $\rb$ is an array of independent
random deviates distributed according to $F_R(r)$;
\item calculation of the arrays $\hb_k= [h_{k1}, h_{k2}, \ldots,
h_{kN}]^T$ with $h_{ki}= {\rm H}_k(x_i)$, $k=0,1,\ldots,N_H$;
\item calculation of the coefficients $\{ a_k \}$ according to
\begin{equation}
a_k=\frac{1}{k!~ N} ~\rb^T ~\hb_k.
\end{equation}
\end{enumerate}

\noindent
The ``optimal'' value for $N_H$ can be determined on the basis of the
value of the parameter $\epsilon$ provided by the criterion
\begin{equation}
\epsilon={\rm DIST}[f_R,f_{r^*}],
\end{equation}
where ${\rm DIST}[.,.]$ is a measure of the distance between $f_R(r)$
and the PDF, $f_{r^*}$, relative to the random deviates
\begin{equation}
\rb^*= \sum_{k=0}^{N_H} a_k {\rm H}_k (\xb).
\end{equation}

Although this approach also presents the problem that $\rho_R(\taub)
\neq \rho_X(\taub)$, here the situation is easier than the method
considered in the previous section. Indeed, because of the orthogonality
of the Hermite polynomials \citep{gra49}, and in particular because of
the so called Kibble-Slepian formula \citep{sle72, dec98}, we have that
\citep{dec98, sak99}
\begin{equation} \label{eq:polcorr}
\rho_R(\taub)= \frac{\sum_{k=1}^{N_H} k! ~a_k^2
~\rho_X^k(\taub)}{\sum_{k=1}^{N_H} k! ~a_k^2}.
\end{equation}
Therefore, $\rho_X(\taub)$ can be obtained by the numerical inversion of
a polynomial function. It is better to recall that, before using it, such 
$\rho_X(\taub)$ must be tested to be a non-negative definite function.

Once $\{ a_k \}$, $N_h$, and $\rho_X(\taub)$ have been determined,
$R(\tb)$ can be obtained by equation (\ref{eq:hexpansion}) that is
implemented in the subroutine {\it Field\_Herm} shown in figure
\ref{fig:code2}. 

\noindent
Some notes on the use of the subroutine {\it HermCoeff}:
\begin{itemize}
\item[-] the input parameters are the length of the arrays $\xb$ and
$\rb$, the PDF $f_R(r)$, the target correlation 
function $\rho_R(\taub)$, and the
parameter $\epsilon$ for the convergence criterion. The output
quantities are the number $N_H$ of terms for the Hermite expansion, the
vector ${\bf a} = \{ a_k \}$ containing the values of the 
coefficients of the expansion,
and the correlation function $\rho_X(\taub)$ of the Gaussian random
field $X(\tb)$ to be used in the subroutine {\it Field\_Herm};
\item[-] typical value for $N$ is several thousands;
\item[-] for the stopping criterion, ${\rm DIST}[.,.] \leq \epsilon$, it
is necessary to choose a distance measure between $f_R(r)$ and the
corresponding approximation $f_{r^*}$. Such a choice, as well as the
value of the parameter $\epsilon$, is very situation dependent. One
possibility is to calculate the difference between the corresponding
histograms. Take note, however, that this method can be troublesome in
case of very skewed distributions. In this case it is advisable to
resort to the methods of PDF estimation that do not make use of binning
of the data as, for example, the kernel and the Johnson empirical
distributions methods \citep{vio94}.
\end{itemize}

\begin{figure}[h]          
\plotone{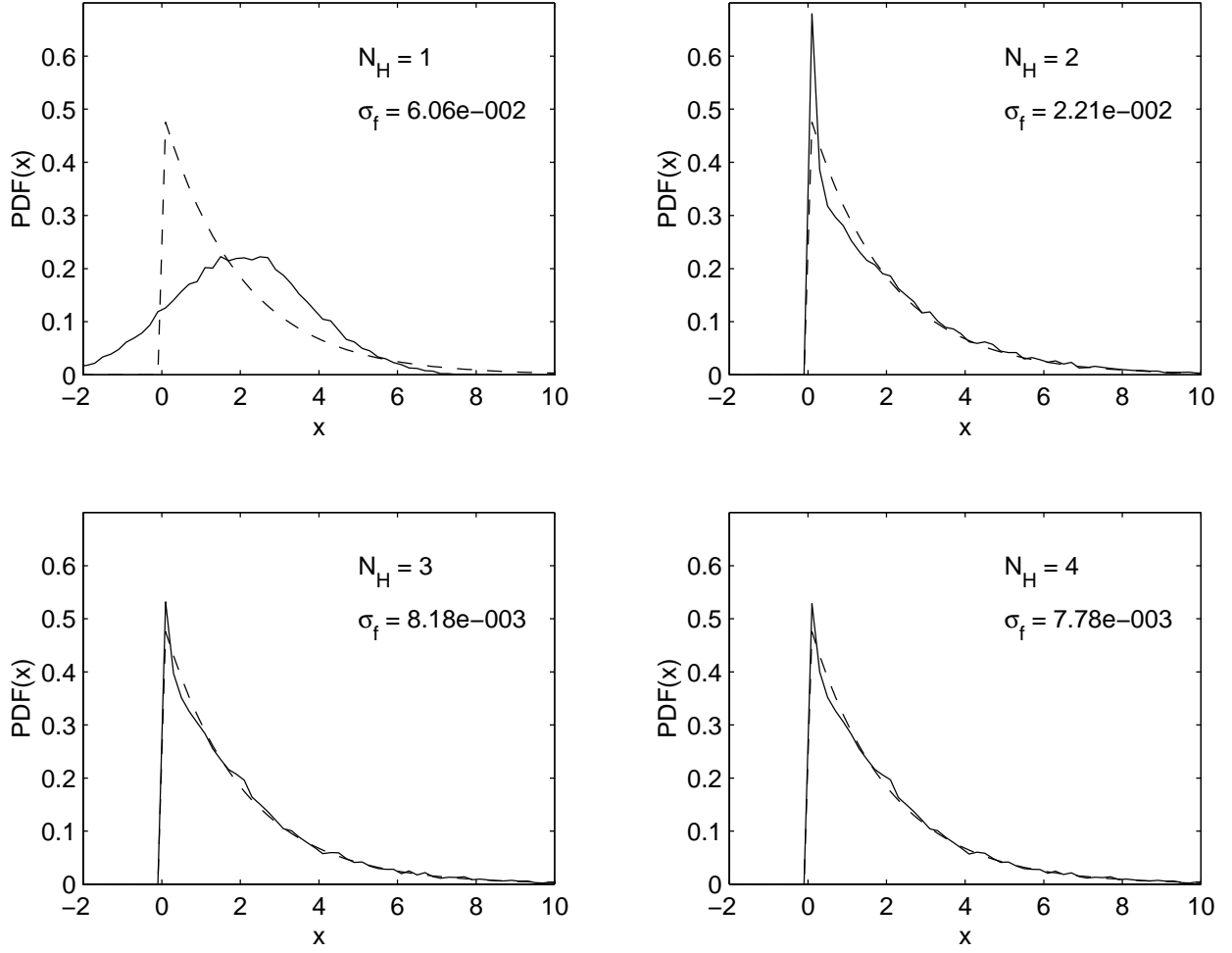}
\caption{\label{fig:hermite} Approximation of the $\chi_2^2$ PDF
obtained through the Hermite polynomial expansion with $N_H$ ranging
from $1$ to $4$. For all the examples $N=16000$. The quantity
$\sigma_f$ is equal to  $\Vert f_R - f_{r^*} \Vert$ (see text), where
$f_{r^*}$ has been obtained via an histogram.}
\end{figure}

\noindent
Some conveniences concerning the algorithm:
\begin{itemize}
\item[-] in case of isotropic random fields, it is possible to work with
an one-dimensional correlation function $\rho_R(\tau)$;
\item[-] once the coefficients ${\bf a}$ of the expansion have been
determined, these can be used for simulating an unlimited number of
random fields $R(\tb)$;
\item[-] the algorithm works also in case of very skewed distributions
(see figure \ref{fig:hermite}).
\end{itemize}

\noindent
One inconvenience in using this algorithm is that $f_R(r)$ is only
approximated and in particular situations this fact can be troublesome.
For example, in case of strictly positive random fields $R(\tb)$, it
could happen that $R^*(\tb)$ presents some negative values. However, if
the approximation is good enough (e.g. only a few values violate the
constraints), the solution to this kind of problem can be very simple
(es. the reflection of the negative values to positive values).

\subsection{Method of Yamazaki \& Shinozuka } \label{sec:iterative}

\begin{figure}[h]          
\plotone{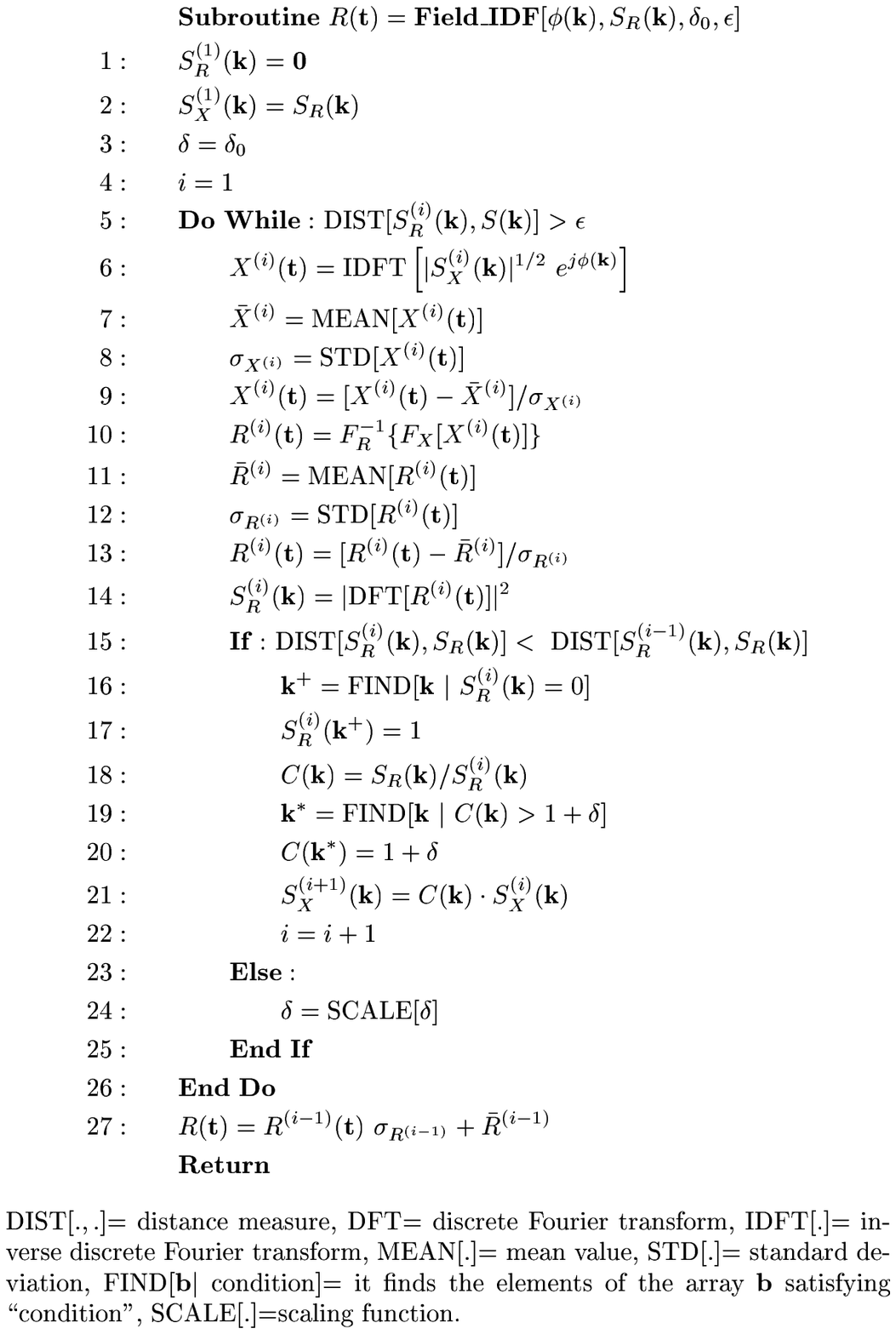}
\caption{\label{fig:code1} Subroutine {\it Field\_IDF}.} 
\end{figure}

Figure \ref{fig:code1} shows the subroutine {\it Field\_IDF}
implementing an algorithm based on an idea by \citet{yam88}. The
rationale behind this code is simple and is based on an iterative
procedure. As explained in section
\ref{sec:notes}, the mapping (\ref{eq:map1}) deforms $\rho_X(\taub)
\rightarrow \rho_R(\taub)$, and consequentely the corresponding power
spectrum $S_X(\kb) \rightarrow S_R(\kb)$, in a complex way. However, if
one applies the transformation (\ref{eq:map1}) to an initial
$X^{(1)}(\tb)$, characterized by a power spectrum $S_{X}^{(1)}(\kb)$ set
equal
to the target
$S_R(\kb)$, it is possible to recover information on the relationship
between $S_{X}^{(1)}(\kb)$ and $S_R(\kb)$
from the power spectrum $S_{R}^{(1)}(\kb)$ of $R^{(1)}(\tb)$. A new
Gaussian
field $X^{(2)}(\tb)$, characterized
by a power spectrum $S_{X}^{(2)}(\kb)$, is then built
with the aim that, after mapping
(\ref{eq:map1}), $S_{R}^{(2)}(\kb)$ is closer to $S_R(\kb)$ than
$S_{R}^{(1)}(\kb)$ . This operation is carried out at line $21$ of the
code where the power spectrum of $X^{(i+1)}(\tb)$ is assumed to be given
by
\begin{equation} \label{eq:siter}
S_X^{(i+1)}(\kb)= C(\kb) S_X^{(i)}(\kb),
\end{equation}
with
\begin{equation}
C(\kb)=\frac{S_R(\kb)}{S_R^{(i)}(\kb)}.
\end{equation}
Through this step, $S_X^{(i)}(\kb)$ is modified according to the
fractional difference, $C(\kb)$, between $S_R(\kb)$ and
$S_R^{(i)}(\kb)$. The entire procedure can be repeated $n$ times until
that $S_{R}^{(n)}(\kb)$ is a good approximation of $S_{R}(\kb)$ .

The code presented in figure \ref{fig:code1} has been modified with
respect to the original version of Yamazaki \& Shinozuka. The main
difference
referes to the implementation of steps $15$, $19$-$20$, $23$-$25$.
The task of these steps is to constrain the range of the permitted
values
for $C(\kb)$. Indeed, strictly speaking, the use of
such a factor is correct only within the hypothesis that the map
(\ref{eq:map1}) is linear.
The consequence is that in many situations $C(\kb)$ appears as a highly
oscillating
function with large extremes even in cases where the target
$S_R(\kb)$, and therefore the starting $S_X^{(1)}(\kb)$, is a smooth
function. Because of this fact, in general the algorithm of Yamazaki \&
Shinozuka
converges only with moderately non-Gaussian fields. Our modification is
based on the idea that, although some values of $C(\kb)$ could be too
large, they are still able to provide indications concerning the
direction of the corrections to make in $S_X^{(i)}(\kb)$, via equation 
(\ref{eq:siter}), for improving the results 
of the iterative process. That suggests the following
procedure:
\begin{enumerate}
\item once $C(\kb)$ was computed, the subset $\kb^*$ of the indices
$\kb$
has to be identified for which $C(\kb^*)$ is larger than
a threshold $1 + \delta$, where $\delta$ is an appropriate value;
\item set $C(\kb^*)=1+\delta$. In this way, it is possible to obtain a
smoothed version of $C(\kb)$ that maintains the original information
on the direction of the correction for each frequency $\kb$;
\item if after this operation it happens that ${\rm
DIST}[S_R^{(i)}(\kb),S_R(\kb)] \geq {\rm
DIST}[S_R^{(i-1)}(\kb),S_R(\kb)]$, where ${\rm DIST}[. ,.]$ indicates a
distance measure between the two arguments, it is necessary to rescale
the parameter $\delta$ according to a prescribed schedule. This point
makes it possible to avoid troublesome oscillations of the algorithm.
\end{enumerate}
From our simulations, it appears that after these modifications the
algorithm converges also in situations where the original method fails.

\noindent
Some notes on the use of the subroutine {\it Field\_IDF}:
\begin{itemize}
\item[-] the first two input quantities are the phase angles, $\phi(\kb)$,
and the power spectrum, $S_X(\kb)$, of a zero-mean, unit-variance, and
Gaussian random field. Here, the important point is that $S_X(\tb)$ is
set equal to the target $S_R(\kb)$.

The third input quantity is the initial value, $\delta_0$, of the
parameter $\delta$. Typically, $\delta_0= 1$-$2$, but such a choice
is not so critical for the final results.

For the fourth input quantity, $\epsilon$, see below;

\item[-] in the stopping criterion, ${\rm DIST}[.,.] \leq \epsilon$, any
measure of distance can be used between $S_R^{(i)}(\kb)$ and $S_R(\kb)$.
An interesting suggestion comes from \citet{pop98} that in their work
use the quantity
\begin{equation}
{\rm DIST}[S_R^{(i)}(\kb), S_R(\kb)]= \frac{\sum_{\kb} \vert
S_R^{(i)}(\kb) - S_R(\kb) \vert}{\sum_{\kb} S_R^{(i)}(\kb)}.
\end{equation}
Typical value for $\epsilon$ are of order of $10^{-2}$-$10^{-3}$;
\item[-] the scaling, ${\rm SCALE}[.]$, of the parameter $\delta$ can
follow any schedule. In our simulations we have halved the value
whenever required by the convergence check.
\end{itemize}

\begin{figure}[h]        
\plotone{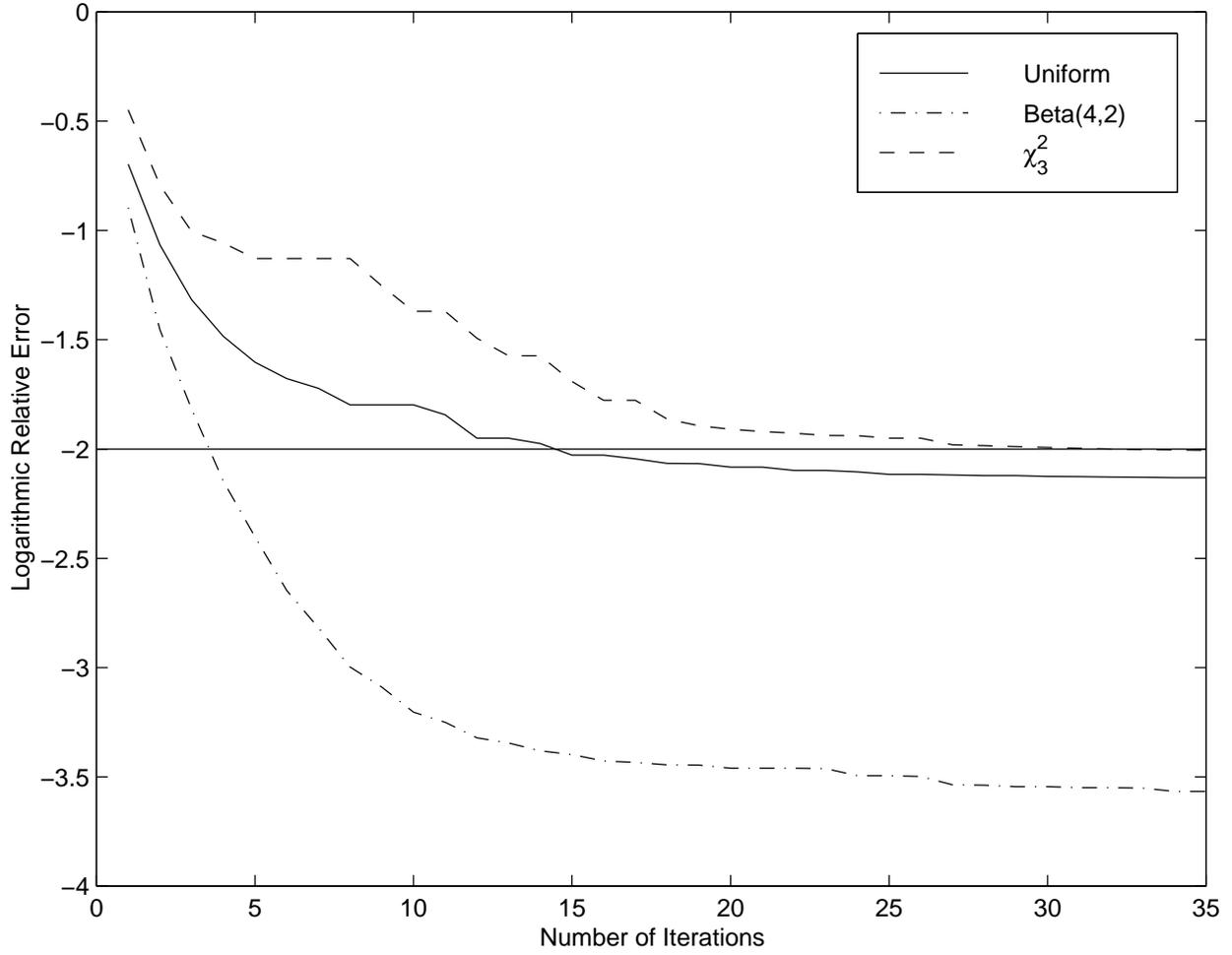}
\caption{\label{fig:fields_err} Convergence rate of the Yamazaki \&
Shinozuka method concerning some type of non-Gaussian fields: Uniform
(skewness= $0$, kurtosis= $-1.2$), $\chi^2_3$ (skewness=
$1.63$, kurtosis= $4$) and Beta(4,2) (skewness= $0.47$, kurtosis
= $0.38$).}
\end{figure}

\begin{figure}[h]       
\plotone{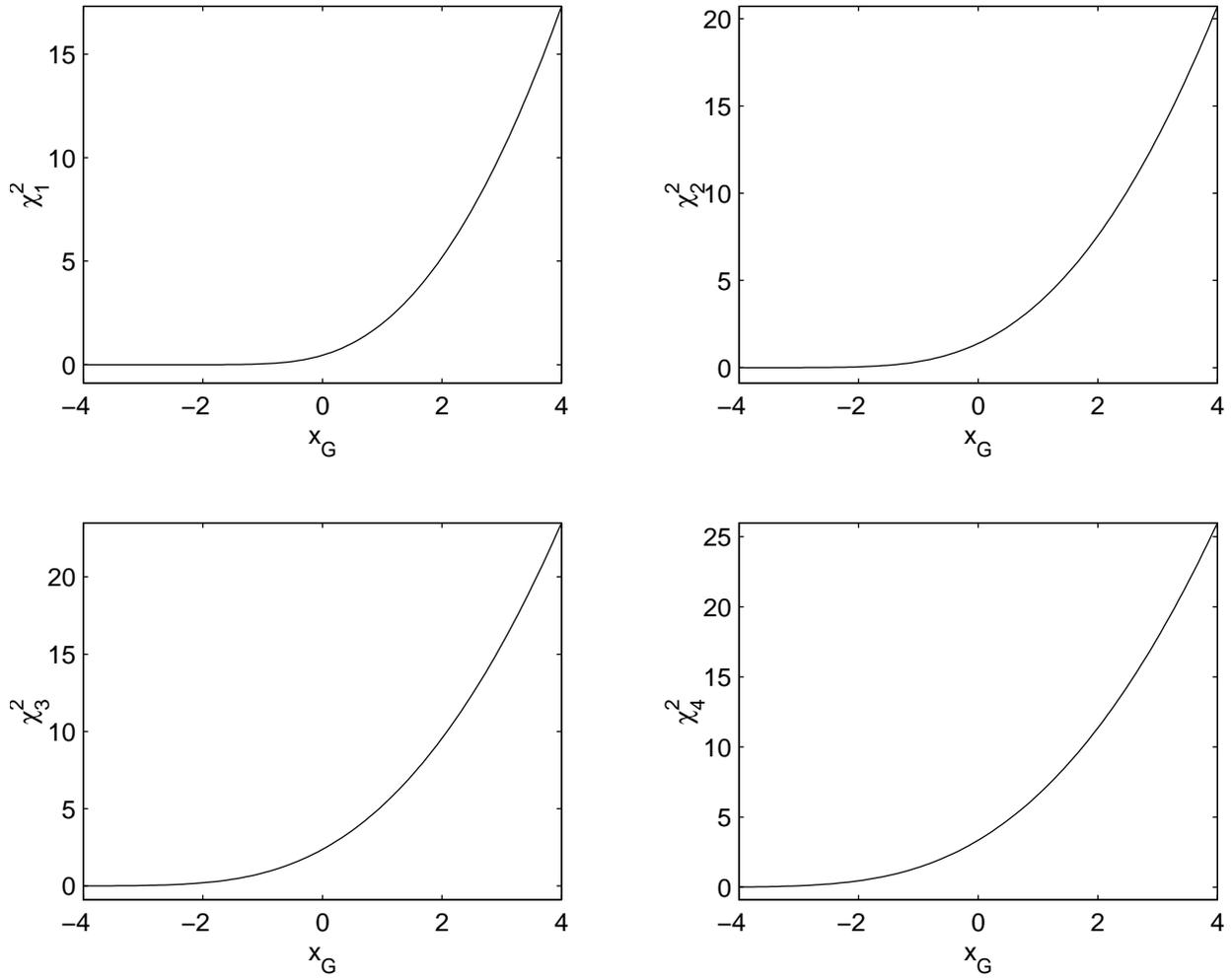}
\caption{\label{fig:Gauss_chi} Mapping (\ref{eq:map1}) concerning the
$\chi^2$ distribution with $1$-$4$ degrees of freedom.}
\end{figure}

In the context of the astronomical applications, some limitations
concerning the algorithm are:
\begin{itemize}
\item[-] the target power spectrum $S_R(\kb)$ must be a smooth function.
That means to work with the expected power spectra of the fields (NB. in
certain engineering applications this is a demanded point). Such
requirement is due to the fact that the generation of $R(\tb)$,
according to the procedure implemented in the algorithm of figure
\ref{fig:code1}, in practice constitutes an optimization problem. Since
the rougher a function, the larger is the corresponding number of
degrees of freedom that must be accounted for by an optimization
procedure, a non-smooth $S_R(\kb)$ will be hardly a solvable problem
with
the present algorithm;
\item[-] although the algorithm is more robust than the original version
of Yamazaki \& Shinozuka, it still presents convergence difficulties in
case of PDFs very different from the Gaussian one (see figure
\ref{fig:fields_err}). In particular, the most serious problems concerns
very skewed distributions. The reason can be understood from the
figures \ref{fig:Gauss_chi}a-d, where the mapping (\ref{eq:map1}) is
presented for four distributions $\chi^2_d$ with $d=1$-$4$. The
first two distributions represent situations that the algorithm is not
able to solve, the third one corresponds to a difficult case, whereas
the last distribution can be easily handled with. It is easy to see that
the most problematic situations concern the mappings where a large
portion of the domain of the Gaussian random variable is projected
onto an almost constant value. The reason is that at the {\it i}-th
iteration of the algorithm, the updated $R^{(i)}(\tb)$ is calculated on

the basis of $X^{(i)}(\tb)$. However, although $X^{(i)}(\tb)=
X^{(i-1)}(\tb) + \Delta^{(i-1)}(\tb)$, it can happen that $R^{(i)}(\tb)
\approx R^{(i-1)}(\tb)$ since $F^{-1}_R \{ F_X [ X^{(i-1)}(\tb) +
\Delta^{(i-1)}(\tb)] \} \approx F^{-1}_R \{ F_X [ X^{(i-1)}(\tb)] \}$.
In this case, the subsequent iterations will be not able to further
improve the result.

Another problem was recently identified by \citet{deo00}:
after the first iteration the field $X^{(i)}(\tb)$ is no
longer strictly Gaussian and therefore the mapping (\ref{eq:map1})
will not give a $R^{(i)}(\tb)$ with the
correct marginal distribution. These authors provide a modified version
of the original algorithm of Yamazaki \& Shinozuka  where, after the
first iteration, $F_X$ is
substituted by an empirical distribution of $X^{(i)}(\tb)$. Actually,
such a method works well also in case of very skewed distributions
and/or non-smooth target power spectra.
Unfortunately, it is very expensive with respect to computational
time which makes its use problematic in practical situations
(e.g. simulation of sizeable random fields);

\item[-] in the case of homogeneous and isotropic {\it N}-dimensional
random fields, it is necessary to work in the {\it N}-dimensional
Fourier domain. Furthermore, the entire procedure must be restarted for
each new simulation.
\end{itemize}

In spite of these problems, the algorithm described in this section
maintains a certain interest since, contrary to other techniques, it can
be easily adapted for the simulation of vector-valued random fields
\citep{pop98}.

\subsection{Fixing the Mean and the Variance}

In all the methods presented in the previous sections, the mean
and the variance of $R(\tb)$ are fixed by $F_R(r)$. However, without
modifying the correlation structure, one can force $R(\tb)$ to have a
given mean $\mu_*$ and a variance $\sigma^2_*$ by means of the
following transformation
\begin{equation}
R_*(\tb)= \mu_* + \frac{R(\tb) - \mu_R}{\sigma_R} ~~ \sigma_*.
\end{equation}

\section{SOME POSSIBLE APPLICATIONS}

As already reported in the Introduction, the numerical simulation of
non-Gaussian random fields can be used in understanding many both
experimental and theoretical physical
problems.

\begin{figure}[h]       
\plotone{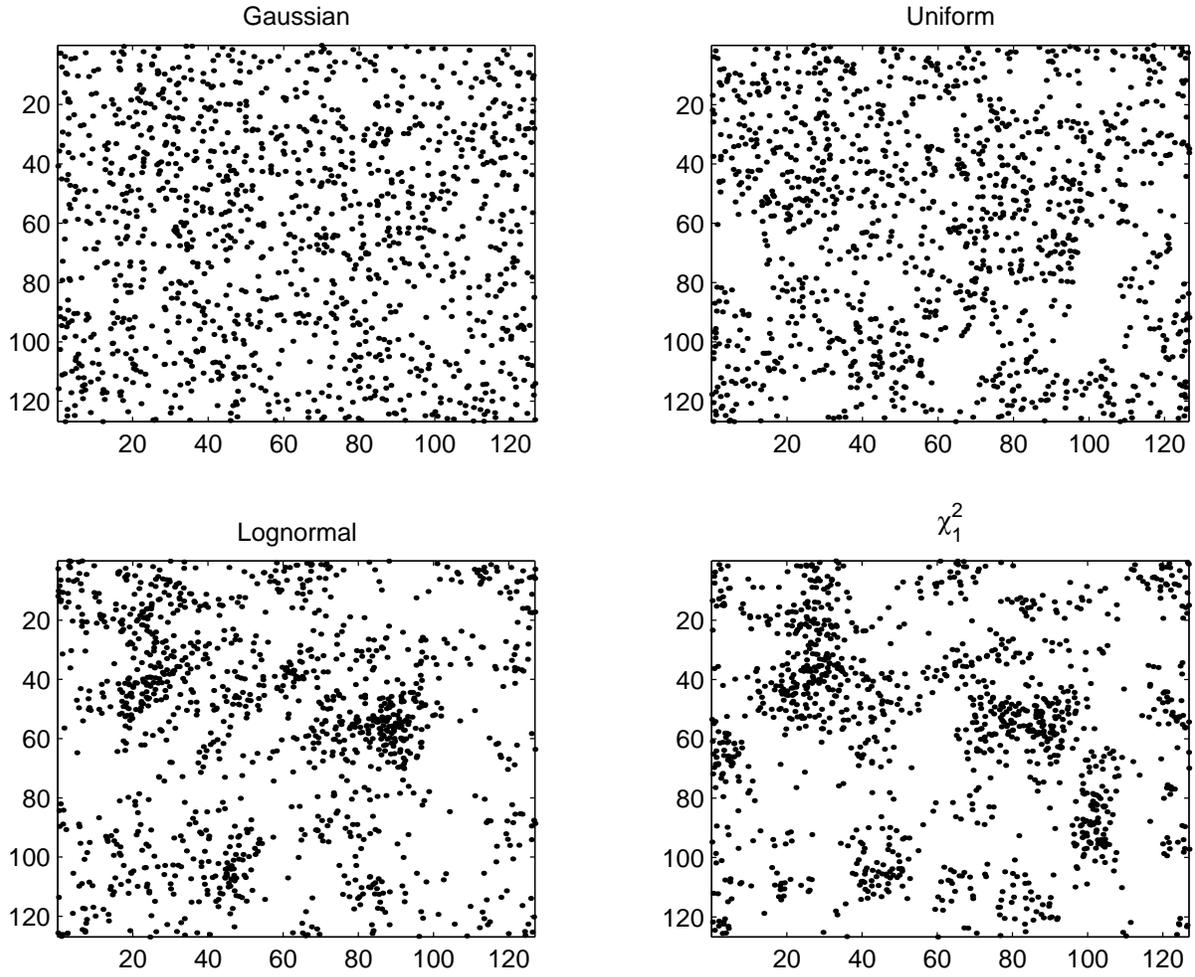}
\caption{\label{fig:points} Point processes obtained from the
non-Gaussian random fields of figure \ref{fig:fields}.}
\end{figure}

Hot topics in Astrophysics and Cosmology,
where the techniques described in this work find quick applications,
can be easily identified as:

\begin{itemize}
\item
the simulation of continuous maps to match the properties of
sky backgrounds. For instance, very deep maps of the extragalactic IR
sky
from space are plagued by the presence of Galactic Cirrus emission
and at very small scales from source confusion;

\item
the generation of non-Gaussian initials conditions for the {\it N}-body
simulations (see figure \ref{fig:points}). Indeed, these conditions can be
obtained by interpreting a non-Gaussian random fields as a density
field. In reality, such approach is not new. However, in the past the
initials conditions were simulated by using only specific distributions
functions as, for example, the Lognormal \citep{mos91,col91} and the
Chi-Square \citep{sco00} ones;

\item
the reconstruction of the missing parts of experimental maps (e.g.
angular distribution of {\it IRAS} galaxies).
Indeed, it is sufficient to transform the non-Gaussian field in a
Gaussian one via the inverse of the mapping (\ref{eq:map1}), to carry
out the desired reconstruction through one among the many techniques
available for the Gaussian case (e.g. \citet{ryb92}), and then to
trasform back the resulting field via the mapping (\ref{eq:map1}). This
techniques is described in detail in \citet{she95} but, again, it is
specialized to the Lognormal case.
\end{itemize}

\section{A PRACTICAL EXAMPLE}

The approach presented in this paper is going to be used to simulate
the FIR sky as it will be observed by the HERSCHEL Satellite \citep{pil01}.
To mock extragalactic catalogues, built on the basis of theoretical modeling
of the expected number of FIR sources, the emission from ``local''
(Galactic and interplanetary) backgrounds has to be added, in order
to reproduce realistic observing conditions (\citet{and00} and \citet{and01}).
Here, as an example, the reproduction of a typical Galactic background,
starting from an observed sky region, is briefly outlined.

\begin{figure}[h]       
\plotone{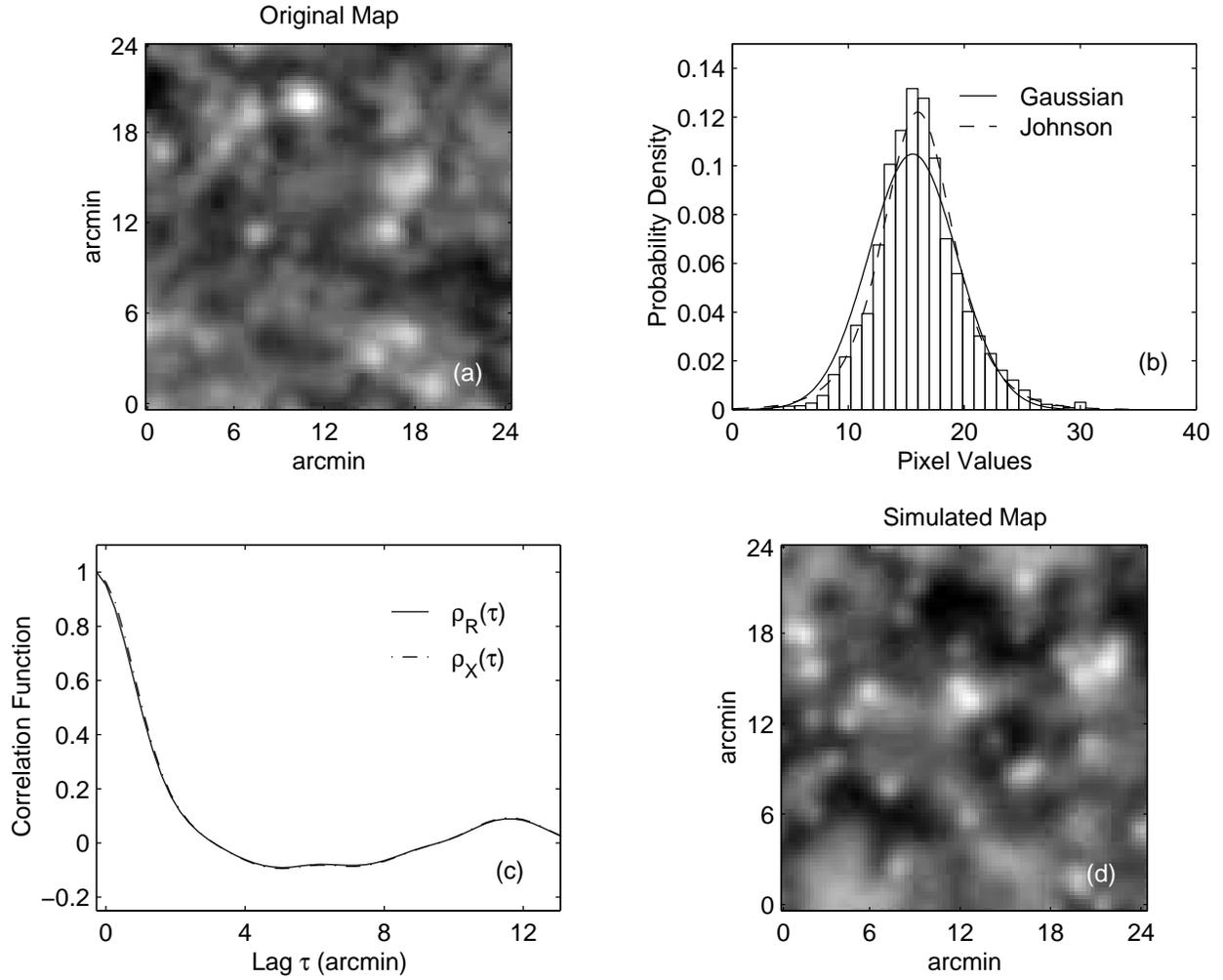}
\caption{\label{fig:simula} a) Original FIR map obtained by the ISO
satellite. b) Classical and Johnson histograms of the values of the map in the previous panel.
The Gaussian PDF is plotted for reference. c) Correlation function of the original map
and correlation function of the Gaussian map used in the numerical experiments (see text).
d) Typical non-Gaussian simulated map.}
\end{figure}

Figure \ref{fig:simula}a shows a sky map, observed by the ISOPHOT camera \citep{lem96}
on board of the ISO Satellite \citep{kes96} at 175$\mu$m, with a
projected size of roughly $24^\prime \times 24^\prime$
\citep{dole01}. The histogram of its values (see figure \ref{fig:simula}b)
shows that the reproduction of such map requires the use of non-gaussian techniques. 
In particular, we have choosen the ``change of variable method'' with the 
following adjustments:
\begin{itemize}
\item[-] the PDF of the pixel values of the original image is
estimated via the Johnson parametric method (for details see 
\citet{vio94}). With such an approach it is possible to build
the mapping (\ref{eq:map1}), as well as its 
inverse, in 
closed form. That results in a much less expensive numerical cost
than in case of the use of the more popular histogram
(see Figure \ref{fig:simula}b);
\item[-] to infer the correlation function $\rho_X(\taub)$, necessary
for the numerical generation of the gaussian random field we prefer not 
to invert equation (\ref{eq:change}). 
Instead, we choose a set of forty-one equidistant values for $\rho_X$ 
in the range $[-0.2,~1.0]$, to determine the corresponding values of 
$\rho_R$ via the numerical integration of equation (\ref{eq:change}), 
and then interpolate the resulting points via a 
cubic-spline. In this way, again, a lot of computational effort is saved. The 
result is shown in figure \ref{fig:simula}c.
\end{itemize}
One of the possible simulations of the original map is shown in
Figure \ref{fig:simula}d.

\section{SUMMARY AND CONCLUSIONS}

In this paper we have considered numerical simulation of
non-Gaussian, scalar random fields $R(\tb)$, with prescribed correlation
structure $\rho_R(\tau)$ and one-dimensional marginal probability
distribution $F_R$, based on the transformation $R(\tb)= g(X(\tb))$
of a Gaussian random field $X(\tb)$. In general, the definition of a
function $g(.)$, able to map the standard random Gaussian variable $x$
in a random variable $r$ with the required $F_R$, is not a difficult
task. Problems are found when the simulated fields {\it has} to have a
desired correlation structure, since in general $\rho_X(\taub)
\neq \rho_R(\taub)$.
The determination
of the appropriate $\rho_X(\taub)$ is achieved using various techniques.
The most effective method is that providing a closed relationship
between
$\rho_X(\taub)$ and $\rho_R(\taub)$. Unfortunately,
this approach can be followed only in a very limited number of cases.
Therefore, in the practical applications, very often it is necessary to
resort to numerical techniques.

Here, we have presented three approaches: the ``change of variable
method'', the ``Hermite expansion method'' and the ``method of Yamazaki
\& Shinozuka". Whenever possible, the first one has to be adopted since,
contrary to the other two, it is able to provide exact results. The only
limitation concerning this method is that typically it requires the
calculation of a large number of double integrals. In certain sitations
that could be computationally too expensive. In this case, it is more
convenient to use the ``Hermite expansion method'' since more robust and
versatile than the ``method of Yamazaki \& Shinozuka". This last method
maintains a certain interest since it can be easily generalized for the
numerical generation of vector-random fields.

\acknowledgements
The authors warmly thank George Deodatis, Sabino Matarrese
and Radu Popescu for helpful discussions on 
the algorithms presented in this work.
P.A. acknowledges support from the Alexander von Humboldt Stiftung
and thanks the IR-Group of the Max-Planck Institut f\"ur
Extraterrestrische
Physik for hospitality. This work was partially financed by
the Italian Space Agency
(ASI) under contract ARS-98-226.

\clearpage

\newpage

\newpage

\newpage

\newpage

\newpage

\newpage

\newpage

\newpage

\end{document}